\documentclass[preprintnumbers,amsmath,amssymb,eqsecnum]{revtex4}

\usepackage{epsf}
\usepackage{graphicx}  
\usepackage{dcolumn}   
\usepackage{bm}        

\begin{document}

\preprint{hep-th/0406057}

\title{Thermodynamics and Stability of Hyperbolic Charged Black Holes }

 \author{Rong-Gen Cai\footnote{e-mail address:
cairg@itp.ac.cn; Rong-Gen$\_$Cai@baylor.edu}}

\address{CASPER, Department of Physics, Baylor University, Waco,
 TX76798-7316, USA \\
  Institute of Theoretical Physics, Chinese
Academy of Sciences,
 P.O. Box 2735, Beijing 100080, China}

\author{Anzhong Wang\footnote{e-mail address:
Anzhong\_Wang@baylor.edu}}

\address{CASPER, Department of Physics, Baylor University, Waco,
 TX76798-7316, USA}

\begin{abstract}
In AdS space the black hole horizon can be a hypersurface with a positive, zero or
negative constant curvature, resulting in different horizon topology. Thermodynamics
and stability of black holes in AdS spaces are quite different for different horizon
curvatures. In this paper we study thermodynamics and stability of hyperbolic charged
black holes with negative constant curvature horizon in the grand canonical ensemble
and canonical ensemble, respectively. They include hyperbolic Reissner-Nordstr\"om black
holes in arbitrary dimensions and hyperbolic black holes in the D=5,4,7 gauged supergravities.
It is found that the associated Gibbs free energies are always negative, which implies that
these black hole solutions are globally stable and black hole phase is dominant in the
grand canonical ensemble, but there is a region in the phase space where black hole is not
locally thermodynamical stable with a negative heat capacity for a given gauge potential.
In the canonical ensemble, the Helmholtz free energies are not always negative and heat
capacities with fixed electric charge are not always positive, which indicates that the
Hawking-Page phase transition may happen and black holes are not always locally thermodynamical
stable.

\end{abstract}
\maketitle

\section{Introduction}
Black holes in anti-de Sitter (AdS) spaces are quite different from black holes in
flat or de Sitter (dS) spaces. In asymptotically flat or dS spaces, the horizon topology
of a four dimensional black hole must be a round sphere $S^2$~\cite{FSW}. In AdS spaces, except
for the positive constant curvature horizon, it is possible to have black holes with zero or
negative constant curvature horizons~\cite{Topology}-\cite{CV}.  Due to the different horizon
structure, the associated thermodynamic properties of black holes are rather different.
Indeed, for the Schwarzschild black holes in AdS spaces, there is a phase transition (named as
Hawking-Page phase transition) between the high temperature black hole phase and low temperature
thermal AdS space~\cite{HP}.  In the AdS/CFT (conformal field theory) correspondence~\cite{AdS},
thermodynamics and phase structure of black holes in AdS spaces can be mapped to those of
dual CFTs. It has been argued by Witten~\cite{Witten} that the Hawking-Page phase transition
of Schwarzschild black holes in AdS spaces can be identified with confinment/unconfirment phase
transition of dual CFTs.  As for the phase transition, it has been indeed found that for the AdS
black holes with zero or negative constant curvature horizon, the Hawking-Page phase transition
does not appear and black hole are always locally thermodynamically stable with positive heat
capacity~\cite{Birm} (see also \cite{Topology,BLP,CS,CEJM}).

The electric charge of AdS black holes can be mapped to the R-charge of supersymmetric dual
CFTs~\cite{Gubser,CS2,Gubser1,Gubser2,Harm}. Therefore it is of equal interest to study
thermodynamics and phase structure of AdS charged black holes. Indeed, it has been shown that
phase structure of AdS charged black holes is rather rich because of the presence of charge,
for example, see \cite{BLP,CS,CEJM,Gubser1}. However, most studies on the AdS charged black holes
are restricted to the case with black hole horizon being a positive constant curvature
hypersurface.  Note that the fact that AdS black holes with zero curvature horizon can be
the large horizon limits of the AdS black holes with positive constant curvature
horizon~\cite{Witten,CEJM,Gubser1}, and that the so-called hyperbolic AdS black holes, namely,
AdS black holes with negative constant curvature horizon, belong to another branch. It is
therefore of necessary to investigate thermodynamics and phase structure of hyperbolic charged
black holes and to see what main differences are among AdS charged black holes with different
characteristic curvature horizons. This is just the goal of the present paper.

When the gauge charge is present, we can consider two different ensembles to study thermodynamics
of charged black holes~\cite{York}. One is the grand canonical ensemble in which the gauge
potential (chemical potential conjugate to the electric charge) is fixed. The other is the
canonical ensemble where the physical electric charge is fixed. In these two ensembles, the
associated free energies are the Gibbs free energy and Helmholtz free energy, respectively.
When the Gibbs free energy or Helmholtz free energy is negative, the black hole phase is
dominant over the thermal AdS background phase. This case implies that the black hole is
globally stable, or say, globally preferred. When the Gibbs (Helmholtz) free energy changes its
sign, the Hawking-Page phase transition appears between the AdS black hole and
thermal AdS background~\cite{HP,Witten}. Here the thermal AdS background corresponds to the
case with vanishing mass parameter and gauge charge, and the AdS background as the vacuum
background is assumed when we calculate the free energies associated with AdS black hole
solutions. Another important property of black hole
thermodynamics is the local stability, which in fact indicates whether a black hole can be
in thermal equilibrium with the thermal bath around the black hole. For example, it is
well-known that a Schwarzschild black hole cannot be in thermal equilibrium within an infinite
thermal bath. This is because the heat capacity of the Schwarzschild is negative. Namely,
when the mass of the black hole increases, the temperature of the black hole decreases.
We will also discuss the local stability of hyperbolic charged black holes, which is determined
 by heat capacities in different ensembles.

The organization of the paper is as follows.  In the next section we start with the
hyperbolic Reissner-Nordstr\"om (RN) black holes in arbitrary dimensions. There we stress how
to rewrite the solution in the isotropic coordinates and how to appropriately choose
parameters to parameterize the solution so that the parameters can cover the whole phase
space, which acts as service to the consequent related discussions for the hyperbolic black
holes in D=5, 4, and 7 dimensional gauged supergravities. For the hyperbolic RN black holes,
we find that the Gibbs free energy is always negative, and the heat capacity with a given
electric potential or a given electric charge is also always positive, but the Helmholtz free energy
can change its sign.  In Sec.~III, IV and V, we discuss the hyperbolic black holes in D=5, 4 and
7 dimensional gauged supergravities, respectively.  Due to the appearance of nontrivial scalar
fields, the thermodynamic properties and stability of the black holes get changed from the case
of hyperbolic RN black holes.  We present our main conclusions and give some
discussions in Sec.~VI.

\section{Hyperbolic RN black holes in arbitrary dimensions}
Let us start from an $(n+2)$-dimensional AdS RN black hole
\begin{equation}
\label{2eq1}
ds^2 =-f(r) dt^2 +f(r)^{-1}dr^2 +r^2 d\Omega_n^2,
\end{equation}
where $d\Omega_n^2$ denotes the line element of an unit
$n$-dimensional sphere and the function $f$ is given by
\begin{equation}
\label{2eq2}
 f(r)= 1-\frac{m}{r^{n-1}} +\frac{\tilde
                  q^2}{r^{2n-2}}+\frac{r^2}{l^2},
\end{equation}
$m$ and $\tilde q$ are the mass parameter and electric charge of
the black hole, respectively. $l^2$ is related to the
$(n+2)$-dimensional cosmological constant via $\Lambda
=-n(n+1)/2l^2$. For the solution (\ref{2eq1}), one has to have
$m>0$, otherwise, the singularity at $r=0$ is naked. When
$m=2|\tilde q|$, the solution turns out to be supersymmetric and
the function $f$ can be written as
\begin{equation}
\label{2eq3}
f =\left(1-\frac{\tilde q}{r^{n-1}}\right)^2
+\frac{r^2}{l^2}.
\end{equation}
Obviously the solution in this case does not describe an AdS black
hole, but a naked singularity. Defining
\begin{equation}
\label{2eq4}
 m = \mu +2q, \ \ \ \tilde q^2 = q(\mu+q), \ \ \ r^{n-1} \to
 r^{n-1} +q,
 \end{equation}
 the solution (\ref{2eq1}) can be rewritten in the isotropic coordinates as
 \begin{equation}
 \label{2eq5}
 ds^2 =-H^{-2} f(r) dt^2 +H^{2/(n-1)}(f(r)^{-1}dr^2
 +r^2d\Omega_n^2),
 \end{equation}
 where
 \begin{equation}
 \label{2eq6}
 f(r)= 1-\frac{\mu}{r^{n-1}} +\frac{r^2}{l^2}H^{2n/(n-1)}, \ \ \
 H=1+\frac{q}{r^{n-1}}.
 \end{equation}
 In this form the supersymmetric limit is obtained by taking $\mu
 \to 0$ while keeping $q$ finite. Furthermore, if introducing
 the so-called boost parameter $\beta$
 \begin{equation}
 \label{2eq7}
 q= \mu \sinh ^2\beta,
 \end{equation}
the harmonic function $H$ can be written as $H = 1+\mu
\sinh^2\beta /r^{n-1}$, the physical electric charge of the
solution $\tilde q = \mu \sinh\beta \cosh\beta$, and the
supersymmetric limit corresponds to the case $\beta \to \infty$
and $\mu \to 0$ while keeping  $q$ finite. The electric potential
of the solution (\ref{2eq5}) is
\begin{equation}
\label{2eq8}
 A_t = \frac{\tilde q}{r^{n-1}+q}dt =\frac{\mu
\sinh\beta\cosh\beta}{r^{n-1} +\mu\sinh^2\beta}dt.
\end{equation}
Note that in the Schwarzschild coordinates (\ref{2eq1}), two
integration constants are $m$ and $\tilde q$, and they are
independent of each other. In the isotropic coordinates
(\ref{2eq5}), they are changed to $\mu$ and $q$, and they could be
related to each other via the relation (\ref{2eq7}). In addition,
in the solution (\ref{2eq5}), $\mu \ge 0$ ( and then $q \ge 0$ via
(\ref{2eq7})). Otherwise, the solution describes a naked
singularity.

 Now we turn to the hyperbolic charged
AdS black hole, whose metric is~\cite{CS}
\begin{equation}
\label{2eq9}
 ds^2= -f(r)dt^2 +f(r)^{-1}dr^2 + r^2 d\Sigma^2_n,
 \end{equation}
 where
  \begin{equation}
  \label{2eq10}
  f(r)= -1 -\frac{m}{r^{n-1}} + \frac{\tilde q^2}{r^{2n-2}}
  +\frac{r^2}{l^2},
  \end{equation}
  $m$ and $\tilde q$ are two integration constants, $d\Sigma_n^2$ stands for the line element for an
 $n$-dimensional hypersurface with negative constant curvature
 $-n(n-1)$.
 One can obtain a closed black hole horizon by acting on the
 hyperbolic space $\Sigma_n$ by discrete subgroup of the isometric
 group of the hyperbolic space, resulting in a higher genus closed
 hypersurface. In four dimensions, the authors of \cite{Klemm} have shown
 that  a magnetic charged hyperbolic solution with $m=0$ and $\tilde
 q^2=l^2/4$ is supersymmetric. In that case, the metric function
 can be written as
 \begin{equation}
 \label{2eq11}
 f= \left(\frac{l}{2r} -\frac{r}{l}\right)^2.
 \end{equation}
 Curiously this is an extremal black hole solution with vanishing
 Hawking temperature, although the mass parameter $m=0$ in this
 solution. In fact, it is now well-known that there are so-called
 ``zero mass'' and ``negative mass'' black holes when the horizon
 is a hyperbolic surface. For example, when $m=0$, the solution (\ref{2eq9})
 still has
 two black hole horizons $r^2_{1,2}=l^2(1\pm \sqrt{1-4\tilde q^2/l^2})/2$
 provided $4\tilde q^2/l^2<1$ in the four dimensional case. In fact,
  for the solution (\ref{2eq10}) the black hole horizon is still present, even the mass
 parameter $m$ is  negative down to
 \begin{equation}
 \label{2eq12}
 m_c = -2r_c^{n-1}\left(1 -\frac{nr_c^2}{(n-1)l^2}\right),
 \end{equation}
 where
 \begin{equation}
 \label{2eq13}
 \frac{r_c^2}{l^2}=\frac{n-1}{n+1}\left( 1 +\frac{\tilde
 q^2}{r_c^{2n-2}}\right).
 \end{equation}
 When $m=m_c$, the solution describes an extremal black hole with
 vanishing Hawking temperature. When $m <m_c$, the
 singularity at $r=0$  becomes naked.  Therefore, for a given
 charge, $r_c$ given in (\ref{2eq13}) is the smallest black hole
 horizon. Despite the horizon structure, so far we have already
 seen the big difference between the AdS RN solution (\ref{2eq1})
 and the hyperbolic charged AdS solution (\ref{2eq9}).

It is instructive to rewrite the hyperbolic black hole solution
(\ref{2eq9}) in terms of isotropic coordinates like in
(\ref{2eq5}). Defining
 \begin{equation}
 \label{2eq14}
 m = \mu - 2 q,  \ \ \ \tilde q^2 =q(\mu -q), \ \ \ r^{n-1} \to
 r^{n-1}+ q,
 \end{equation}
 we then have
 \begin{equation}
 \label{2eq15}
ds^2= -H^{-2} f(r) dt^2 +H^{2/(n-1)}(f(r)^{-1}dr^2
 +r^2d\Sigma_n^2),
 \end{equation}
 where
 \begin{equation}
 \label{2eq16}
 f(r)= -1-\frac{\mu}{r^{n-1}} +\frac{r^2}{l^2}H^{2n/(n-1)}, \ \ \
 H=1+\frac{q}{r^{n-1}}.
 \end{equation}
 In these coordinates, we can see that both signs (positive and negative)
 of $\mu$ are allowed to have black hole horizon. (1) when $\mu >0$, one has to
 have  $0<q<\mu$ in order to keep the electric charge
 squared $\tilde q^2$ positive. (2) When $\mu <0$, instead one has $ \mu <q <0$.
  Combining these two cases, we find that it is appropriate to introduce
  a parameter $\alpha$ to parameterize the charge $q$ as
 \begin{equation}
 \label{2eq17}
 q = \mu \sin^2\alpha.
 \end{equation}
One then  has $H = 1+\mu \sin^2\alpha /r^{n-1}$, the
physical electric charge of the solution
 $\tilde q = \mu \sin\alpha \cos\alpha$, and the corresponding
 electric potential
 \begin{equation}
 \label{2eq18}
 A_t = \frac{\mu \sin\alpha \cos\alpha}{r^{n-1} +\mu
 \sin^2\alpha}dt=\frac{\tilde q}{r^{n-1} +q}dt.
 \end{equation}
 Note that when $ q<0$, the original singularity ar $r=0$ in the
 coordinates (\ref{2eq9}) is moved to $r=|q|$ in the coordinates
 (\ref{2eq15}).  Furthermore, the supersymmetric solution (\ref{2eq11}) is
achieved by taking $\sin^2\alpha =1/2$ in the coordinates
(\ref{2eq15}).

 Note that $m$ and $\tilde q$ in (\ref{2eq10}) are two
 integration constants which are independently variable in the range
$(-\infty, \infty)$. In the isotropic coordinates (\ref{2eq15}),
these two integration constants are mapped to $\mu $ and $q$. From
(\ref{2eq17}), however, one can see $0 \le |q| \le |\mu|$. Namely,
given a $\mu$, one cannot get an arbitrary large $q$.  Therefore,
$\mu$ and $q$ through (\ref{2eq14}) does not cover the whole phase
space given by $m$ and $\tilde q$, although in that case the
expressions in (\ref{2eq16}) have very similar forms as those
(\ref{2eq6}) of the AdS RN black hole. To remedy this difficulty,
we find that the parameters $m$ and $q$ are good to do
that. In this case, one has
\begin{eqnarray}
\label{2eq19}
 && \tilde q^2 = q(m+q); \nonumber \\
 &&  f(r)= -1-\frac{m+2q}{r^{n-1}} +\frac{r^2}{l^2}H^{2n/(n-1)}, \ \ \
 H=1+\frac{q}{r^{n-1}}.
 \end{eqnarray}
 Furthermore, instead of (\ref{2eq17}), we can introduce a
 parameter $\beta$ to parameterize the charge $q$ as $q =m
 \sinh^2\beta$ with $0\le \beta <\infty$. The physical electric charge becomes
 $\tilde q^2 = m^2\sinh^2\beta \cosh^2\beta$. Thus, we see that $m$ and $q$
 indeed cover the whole phase space as $m$ and $\tilde q$ in
 (\ref{2eq9}).

 However, it turns out that it is more convenient to use the
 Schwarzschild coordinates (\ref{2eq9}) to discuss the thermodynamic
 properties and stability of the hyperbolic black hole.
 In terms of horizon radius $r_+$, determined by $f(r)|_{r=r_+}=0$,
 for the hyperbolic black hole (\ref{2eq9}), the mass $M$,
 Hawking temperature $T$, entropy $S$ and chemical potential
 $\phi$ which is conjugate to the electric charge $\tilde q$
 are found to be
 \begin{eqnarray}
 \label{2eq20}
&& M = \frac{n V_n}{16\pi G}m =\frac{nV_nr_+^{n-1}}{16\pi G}\left (-1
  +\frac{\tilde q^2}{r_+^{2n-2}}+\frac{r_+^2}{l^2}\right)  , \\
&& T =\frac{(n-1)}{4\pi r_+}\left(-1 -\frac{\tilde
q^2}{r_+^{2n-2}}
 +\frac{n+1}{n-1}\frac{r_+^2}{l^2}\right), \\
&& S = \frac{V_n}{4 G}r_+^n, \\
\label{2eq23}
 && \phi =\frac{nV_n}{16\pi G} \frac{2\tilde
q}{r_+^{n-1}},
\end{eqnarray}
where $V_n$ is the volume of the unit hyperbolic surface
$\Sigma_n$. It can be checked that these quantities obey the first
law of black hole thermodynamics
\begin{equation}
\label{2eq24}
 dM = TdS +\phi d\tilde q.
\end{equation}
According to the AdS/CFT correspondence, the dual CFT resides on
the boundary of the AdS space, whose metric can be determined from
the bulk metric up to a conformal factor,
\begin{equation}
\label{2eq25}
 ds^2 = -dt^2 +l^2 d\Sigma_n^2.
 \end{equation}
 The thermodynamics of the hyperbolic black hole can be mapped to
 that of the dual CFT residing on (\ref{2eq25}). From these
 thermodynamic quantities (\ref{2eq20})-(\ref{2eq23}), we find
 that they satisfy the Cardy-Verlinde-like formula
 \begin{equation}
 \label{2eq27}
 S =\frac{2\pi l}{n}\sqrt{|E_c|(2(M-E_q)-E_c)},
 \end{equation}
 where $E_c=-nV_nr_+^{n-1}/8\pi G$ is the non-extensive part of energy,
 namely the Casimir energy, and $E_q=\phi \tilde q/2$ is the
 energy of electromagnetic field outside the black hole. Thus we
 generalize the discussion made for neutral hyperbolic black
 hole~\cite{CV} to the R-charged case.

  Next we discuss thermodynamic stability and phase structure of the solution.
   In the grand canonical
ensemble, we will fix the chemical potential conjugate to the
electric charge~\cite{York}. The Euclidean action ${\cal I}$ of
the black hole solution has a relation to the Gibbs free energy
${\cal G}$ through ${\cal I}={\cal G}/T$~\cite{York,BLP}.
Therefore to see the global stability of the hyperbolic black hole
as a thermodynamic system, we can calculate the Gibbs free energy,
which is ${\cal G}=M-TS -\phi \tilde q$ by definition,
\begin{equation}
\label{2eq26}
 {\cal G} =r_+^{n-1}\left( -1 -\left(\frac{8\pi G}{nV_n}\phi\right)^2
  -\frac{r_+^2}{l^2}\right) \frac{V_n}{16\pi G}.
 \end{equation}
 This Gibbs free energy is always negative, the hyperbolic black
 hole phase is therefore globally preferred, and the Hawking-Page transition
 will not happen here. Here some remarks are in order. Although the relation
 ${\cal G}= T {\cal I}$ between the Gibbs free energy ${\cal G}$ and the (reduced)
  Euclidean action ${\cal I}$ of black holes always holds (see Refs.~\cite{York,BLP}
  and references therein), some subtleties exist, which concern the understanding of
  the Euclidean action and free energy. When using the relation
 ${\cal G}= T {\cal I}$ and the definition of free energy ${\cal G}= M-TS-\phi\tilde q$ here,
  we have taken the vacuum with $m=0$ and $\tilde q=0$ as the
 reference background, as in \cite{York,BLP}. In contrast to the cases of
 $k=0$ and $k=-1$, there the vacuum backgrounds are pure AdS
 spaces, not a black hole, the vacuum background of $k=-1$ is a
 back hole with horizon $r_+=l$, whose free energy is ${\cal
 G}_0=-l^{n-1}V_n/(8 \pi G)$. Subtracting this contribution, the Euclidean
 action should be understood as ${\cal I }= ({\cal G}-{\cal
 G}_0)T$. The action is always negative provided $r_+ >l$, and is
 zero for the vacuum background. On the other hand, if taking the
 extremal black hole (\ref{2eq12}) as the reference background, as
 in \cite{Birm}, the black hole mass $M$ should be replaced by
 $M-M_c$, where $M_c=nV_nm_c/16\pi G$ with $m_c$ given by (\ref{2eq12}).
 It is easy to show that the Gibbs free energy ${\cal
 G}=(M-M_c)-TS-\phi \tilde q$ is always negative and approaches to
 zero for the extremal black hole. This is a natural generalization
 of the neutral hyperbolic black hole case discussed in
 \cite{Birm}.

 In the grand canonical ensemble, the local stability  condition of a thermodynamic system
 can be determined by the negative definiteness of the Hessian
 matrix. The Hessian matrix is arranged as the second derivatives
 of entropy with some extensive quantities. In our case, the
 extensive quantities are the black hole mass and charge. This
 approach is equivalent to calculating the heat capacity with the
 fixed chemical potential.
  The heat capacity of the hyperbolic black hole with a fixed electric
  potential (chemical potential) $\phi$ is given by
 \begin{eqnarray}
 \label{2eq28}
 C_{\phi} &\equiv & T\left(\frac{\partial S}{\partial
 T}\right)_{\phi} \nonumber \\
    &=&\frac{n\pi
    V_nTr_+^{n+1}}{(n-1)G}\left(1+\frac{n+1}{n-1}\frac{r_+^2}{l^2}
     +2\left(\frac{8\pi G}{nV_n}\phi\right)^2\right)^{-1}.
\end{eqnarray}
One immediately sees that the heat capacity is always positive,
which indicates the stability of the black hole in all range of
temperature.

In the canonical ensemble, we will fix the electric charge of the
system~\cite{York}. In this case, the Euclidean action of the
solution will have a relation to the Helmholtz free energy as
${\cal I}={\cal F}/T$~\cite{BLP,York}, here ${\cal F}= M-TS$. Then
for the present hyperbolic black holes, we have
\begin{equation}
\label{2eq29}
 {\cal F}=r_+^{n-1} \left(-1-\frac{r_+^2}{l^2}+(2n-1)\frac{\tilde
 q^2}{r_+^{2n-2}}\right)\frac{V_n}{16\pi G}.
\end{equation}
This indicates that when $\tilde q^2 > \tilde q_c^2$, where
\begin{equation}
\label{2eq30}
  \tilde q^2_c = \frac{r_+^{2n-2}}{2n-1} \left(
  1+\frac{r_+^2}{l^2}\right),
  \end{equation}
  the free energy is positive; otherwise it is negative. This
  implies that in the canonical ensemble with a fixed charge, when
  the temperature $T<T_c$, where
  \begin{equation}
  T_c =\frac{(n-1)}{4\pi r_+}\left(-1 -\frac{\tilde
q^2_c}{r_+^{2n-2}}
 +\frac{n+1}{n-1}\frac{r_+^2}{l^2}\right),
 \end{equation}
 the black hole phase is not globally preferred, instead the
 thermal vacuum background is preferred. When $T>T_c$, the hyperbolic
 black hole phase is globally preferred.  Across the temperature $T_c$,
 a Hawking-Page phase transition happens. In Fig.~1, we plot the
 physical charge squared versus horizon radius. The curve cross
 the horizon axe at $l/\sqrt{2} \approx 0.71l$ denotes the extremal
 black holes with vanishing temperature. The region on the right
 side of the curve is physically allowed one with nonvanishing
 temperature. The other curve denotes the black hole solution
 having zero Helmholtz free energy. On the left (right) side of the
 curve the free energy is positive (negative). From the figure one
 can see that the smallest horizon of black hole is $l/\sqrt{2}$.
 The region with negative $\tilde q^2$ should be excluded in the
 physical phase space.
\begin{figure}[ht]
\includegraphics[totalheight=1.7in]{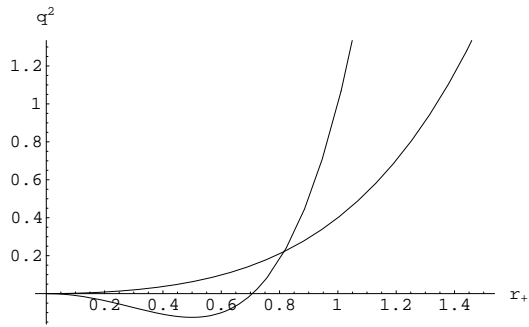}
 \caption{The curves denote the charge squared $\tilde q^2/l^4$ versus the
 horizon radius $r_+/l$ for the case of $n=3$. In the region on the common right of
 the two curves the black hole phase is globally stable in the
 canonical ensemble. The curve across the horizon axe at $l/\sqrt{2} \approx 0.71l$
 denotes the extremal black holes with vanishing temperature.}
\end{figure}

As in the grand canonical ensemble, the Helmholtz free energy and
then the Euclidean action are  also dependent of the choice of the
reference background. It should be
 emphasized here that as we calculate the Helmholtz free energy
 (\ref{2eq29}), the background is taken as the solution
 (\ref{2eq9}) with $m=\tilde q =0$, namely the so-called zero mass
 black hole. This might not be appropriate since the background
 has a vanishing electric charge, which does not satisfy the
 requirement of canonical ensemble. One of the choices as the
 background vacuums is to take the solution with $m=0$ and $\tilde q^2=
 \tilde q_c^2$. In this case, the background is also a black hole
 spacetime. The other is to take the solution (\ref{2eq9}) with
 mass $m=m_c$ in (\ref{2eq12}). In this case, the background is an
 extremal black hole. In either case, we have to recalculate the
 Euclidean action in order to see the globally stability. However,
 we expect that the main conclusions remain valid. We do not to stress here
 the difference arising from different vacuum backgrounds. For related
 discussion see \cite{CEJM}.

The heat capacity of black hole with a fixed electric charge is
found to be
\begin{eqnarray}
\label{2eq31}
 C_{\tilde q} &\equiv & T\left (\frac{\partial
S}{\partial
T}\right)_{\tilde q} \nonumber \\
  &=&\frac{n\pi
    V_nTr_+^{n+1}}{(n-1)G}\left(1+\frac{n+1}{n-1}\frac{r_+^2}{l^2}
     +\frac{(2n-1)\tilde q^2}{r_+^{2n-2}}\right)^{-1}.
\end{eqnarray}
This heat capacity, as the case in the grand canonical ensemble, is
always positive. As a result, the black hole is always locally
thermodynamically stable.

\section{Hyperbolic black holes in $D=5$ gauged supergravity}

The R-charged black holes in D=5, N=8 gauged supergravity have
been found in \cite{BCS} as a special case (STU model) of the
solutions of equations of motion of the D=5, N=2 gauged
supergravity (see also \cite{Duff}). The black hole solution has
the form
\begin{eqnarray}
\label{3eq1}
 &&  ds^2 = -(H_1H_2H_3)^{-2/3}fdt^2
              +(H_1H_2H_3)^{1/3}(fdr^2 +r^2 d\Omega_{3,k}^2), \\
  \label{3eq2}
 && X_i = H_i^{-1} (H_1H_2H_3)^{1/3}, \ \ \ A^i
 =\sqrt{k}(1-H_i^{-1})\coth\beta_idt,
\end{eqnarray}
where
\begin{equation}
\label{3eq3}
 f = k-\frac{\mu}{r^2}+r^2l^{-2}H_1H_2H_3, \ \ \ H_i =
1 +\frac{\mu \sinh^2\beta_i}{kr^2}, \ \ \ i =1,2,3
\end{equation}
where $k$ can be $1$, $0$ or $-1$, corresponding to the foliating
surfaces of transverse space being $S^3$, $T^3$ or $H^3$ with unit
metric $d\Omega_{3,k}^2$, $\mu$ is the mass parameter and
$\beta_i$ are three boost parameters.  When $k=1$ or $k=0$, the
supersymmetric limit of the solution (\ref{3eq1}) is obtained via
taking $\beta_i \to \infty$ and $\mu \to 0$, while keeping $\mu
\sinh^2\beta_i$ finite. Note that when $k=0$, one has to rescale
$\sinh^2\beta_i \to k\sinh^2\beta_i$, and then send $k$ to zero,
resulting in the gauge potential $A^i = (1-H^{-1}_i)
dt/\sinh\beta_i$. Therefore, the supersymmetric solution has
vanishing gauge potential in the case of $k=0$.  The
thermodynamics of cases $k=1$ and $k=0$ has been discussed
in~\cite{Gubser1}. Here we focus on the case $k=-1$, namely the
case where the horizon surface is a hyperbolic 3-space with
constant negative curvature. In this case, the supersymmetric
limit of the solution cannot be obtained as the cases of $k=1$ and
$k=0$. So far it has not yet been clear whether the solution has
the supersymmetric limit, or under what condition the solution is
supersymmetric. In particular, let us note that in the solution
(\ref{3eq1}), if one defines $q_i = \mu\sinh^2\beta_i/k$, the
physical electric charges are
\begin{equation}
\tilde q_i =\sqrt{k} \mu \sinh\beta_i\cosh\beta_i.
\end{equation}
When $k=-1$, one can see from (\ref{3eq2}) that the electric
potentials $A^i$ are pure imaginary if keeping $\beta_i$ real.
This indicates that the parametrization used in (\ref{3eq2}) and
(\ref{3eq3}) is not suitable for the case of $k=-1$. Note that the
solution is written down in the isotropic coordinates and it will
reduce to the D=5 hyperbolic AdS RN black hole in (\ref{2eq9}) as
all three charges are equal in (\ref{3eq1}). According to the
experience in the previous section, we find that the suitable
parametrization can be obtained by the replacement
\begin{equation}
\beta_i  \to  \i \alpha_i,
\end{equation}
when $k=-1$.  Here $\i$ denotes the unit pure imaginary number.
Upon this replacement, we have
\begin{equation}
H_i = 1+\frac{q_i}{r^2}, \ \ \ A^i =\frac{\tilde q_i}{r^2 +q_i}dt,
\end{equation}
where
\begin{equation}
\label{3eq7}
 q_i = \mu \sin^2\alpha_i, \ \ \ \tilde q_i^2 = q_i
(\mu -q_i).
\end{equation}
As stressed in the previous section, in fact, this
parametrization is also unsuitable due to the restriction $0\le
|q| \le |\mu|$. We find, however, that when the solution has only
an electric  charge parameter, which includes three cases: (i)
three charges are equal, $q_1=q_2=q_3=q$; (ii) two charges are
equal, the third vanishes, say, $q_1=q_2=q$ and $q_3=0$; (iii) one
charge does not vanish, say $q_1=q$, and the other two vanish, one can
remedy the difficulty  $|q_i| \le |\mu$ in (\ref{3eq7}).
Introducing
\begin{equation}
\mu = m + 2 q, \ \ \ \tilde q^2 = q (m+q),
\end{equation}
we  then have
\begin{equation}
f = -1 -\frac{m+2q}{r^2} +r^2l^{-2} H_1H_2H_3, \ \ \ H_i = 1
+\frac{q}{r^2},
\end{equation}
where $H_1H_2H_3$ depends on the number of nonvanishing charges. In
that case, $m$ and $q$ become two parameters varying independently
in the range $(-\infty, \infty)$, and $q$ can be parameterized via
$q =m \sinh^2\beta$.  Note that in this parametrization, one has
$q>0 (<0)$ if $m >0 (<0)$.

Using the background subtraction approach~\cite{BJ}, the mass of
the hyperbolic black hole can be calculated following the case of
$k=1$ \cite{BCS}. Rewriting the solution in terms of Schwarzschild
coordinates
\begin{equation}
ds^2 = -e^{-2V}dt^2 + e^{2W}dR^2 +R^2 d\Omega_{3,-1}^2,
\end{equation}
where $R^2= r^2 (H_1H_2H_3)^{1/3}$ and
\begin{equation}
e^{-2V} = f (H_1H_2H_3)^{-2/3}, \ \ \ e^{2W} = f^{-1}
(H_1H_2H_3)^{1/3} \left (\frac{\partial r}{\partial R}\right)^2,
\end{equation}
one can find the mass of the solution using the formula
\begin{equation}
\label{3eq12}
M =-\frac{1}{8\pi G} \int_{\partial {\cal M}}N (K-K_0),
\end{equation}
where $N=e^{-V}$ is the norm of the time-like Killing vector and
$K$ is the extrinsic curvature of induced boundary. $K_0$ is the
same as $K$, but for the vacuum background, namely the solution
with $\mu=0$ and $q_i=0$. After calculation, it turns out that the mass of the
hyperbolic black hole is
\begin{equation}
M = \frac{V_3}{8\pi G}\left(\frac{3}{2}\mu-\sum_iq_i\right).
\end{equation}
It is interesting to compare with the cases of $k=1$ and $k=0$, where
the mass is $M = V_3 (3 \mu/2 +\sum_iq_i)/8\pi G$ for $k=1$, while
$M = 3V_3 \mu/16\pi G$ for $k=0$. Here $V_3$ is the volume of
$\Omega_{3,k}$.

The event horizon of the hyperbolic black hole is determined by
$f(r)|_{r=r_+}=0$. In terms of the horizon radius, the mass
parameter $\mu$ can be expressed as
\begin{equation}
\mu = r_+^2(-1+\Pi_{i} \rho_i^2/r_+^4l^2), \ \ \ \rho_i^2 = r_+^2
+q_i.
\end{equation}
The Hawking temperature $T$, entropy $S$ and chemical potential
$\phi_i$ conjugate to the electric charge $\tilde q_i$ are
\begin{eqnarray}
&& T =\frac{r_+^2}{2\pi \Pi_i\rho_i}\left(-1 -\frac{\Pi_i
\rho_i^2}{l^2r_+^4}
+\frac{\Pi_i\rho_i^2}{l^2r_+^2}\sum_i\frac{1}{\rho_i^2}\right),\\
 && S =\frac{V_3}{4G}\Pi_i \rho_i, \\
&& \phi_i = \frac{ V_3}{8\pi G}\frac{\tilde q_i}{\rho_i^2},
\end{eqnarray}
respectively. In terms of the horizon radius $r_+$, the expression
of the mass of black hole is
\begin{equation}
M =\frac{3V_3}{16\pi G} \left(r_+^2
+\frac{\Pi_i\rho_i^2}{l^2r_+^2}-\frac{2}{3}\sum_i\rho_i^2\right).
\end{equation}
It is easy to show that the first law of thermodynamics holds
here, $dM =TdS +\sum_i \phi_i d\tilde q_i$. For this black hole,
the Gibbs free energy, defined as ${\cal G}= M-TS -\sum_i \phi_i
\tilde q$, is found to be
\begin{equation}
{\cal G}=\frac{V_3}{16\pi G}\left(-r_+^2
-\frac{\Pi_i\rho_i^2}{l^2r_+^2}\right),
\end{equation}
which is always negative. The negative definiteness indicates that
the Hawking-Page phase transition will not occur. The black hole
phase always dominates in the whole range of temperature and
chemical potential. Here it also should be stressed that the same
discussions made in the previous section for the case of
hyperbolic RN black holes are applicable. The reference background
is a five dimensional ``zero mass" black hole with $\mu=q_i=0$,
whose free energy is ${\cal G}_0= -l^2V_3/8\pi G$. The reduced
action should be understood as ${\cal I}=({\cal G}-{\cal G}_0)/T$,
which is negative provided $r_+>l$. If taking the extremal black
hole with vanishing Hawking temperature as the vacuum background,
the black hole mass $M$ should be replaced by $M-M_c$, once again.
In this case, the resulting Euclidean action is always negative.

The Helmholtz free energy, defined as ${\cal F}=M-TS$, is
\begin{equation}
{\cal F}=\frac{V_3}{16\pi G} \left( 5r_+^2
+5\frac{\Pi_i\rho_i^2}{l^2 r_+^2}-2\sum_i\rho_i^2
-2\frac{\Pi_i\rho_i^2}{l^2}\sum_i\frac{1}{\rho_i^2}\right).
\end{equation}
This free energy can be negative or positive. To see clearly this,
let us consider a single charge case.
Namely, one of three charges of the solution, say, $q_1=q \ne 0$,
and other two $q_2=q_3=0$. In this case, the equation $f(r)=0$ has
the following roots
\begin{equation}
r^2_{1,2}= \frac{l^2}{2}\left( \left(1-\frac{q}{l^2}\right)
 \pm
 \sqrt{\left(1-\frac{q}{l^2}\right)^2+4\frac{\mu}{l^2}}\right).
 \end{equation}
From the above, we see some interesting features of the black
hole. (1) When  $q<l^2$, the hyperbolic black hole has two
horizons if $- l^2(1-q/l^2)^2/4<\mu <0$, while it has only one
horizon if $\mu>0$.  When $\mu = -l^2(1-q/l^2)^2/4$, these two
horizons coincides with each other. This case corresponds to the
extremal black holes with  vanishing Hawking temperature.
 (2) When $q>l^2$, the solution has one black hole horizon only if
 $\mu>0$. (3) When $q=l^2$, the black hole has a horizon with
 $r_+^2=l\sqrt{\mu}$. (4) When $q=0$, the solution goes back to
 the case of neutral hyperbolic black hole. (5) When $\mu=0$, the
 black hole has also two horizons: one of them $r_+=0$, the other
 is $r_+^2= l^2 -q$ if $q<l^2$.

 In the single charge case, the Hawking temperature is
 \begin{equation}
 T= \frac{1}{2\pi \rho}\left (-1 +\frac{\rho^2}{l^2}
 +\frac{r_+^2}{l^2}\right), \ \ \ \rho^2 =r_+^2 +q,
 \end{equation}
and the Helmholtz free energy reduces to
 \begin{equation}
 {\cal F}= \frac{V_3}{16\pi G}\left(-r_+^2 - 2 q
 -\frac{r_+^4}{l^2} +\frac{r_+^2}{l^2}q\right).
 \end{equation}
 Therefore, when $r_+^2/l^2 \le 2$, the free energy is always negative,
 while $q < r^2(1+r^2/l^2)/(r^2/l^2-2)$ as $r^2_+ /l^2>2$. In Fig.~2 the
 region with negative free energy is plotted between the two
 curves. The curve starting at $q=1$ and $r_+=0$ and ending at $q=0$ and
 $r_+/l=1/\sqrt{2} \approx 0.71$  denotes extremal black holes. So the
 region below this curve should be excluded in the physical phase space. The above
 curve in Fig.~2 denotes the black holes with zero free energy.
 Therefore, a Hawking-Page phase transition happens across that
 curve. Note that the free energy depends
 on the choice of vacuum background. Here the vacuum background
 is taken as the ``zero mass" black hole with $\mu=q=0$. Other
 choices are of course interest to further investigate, as the
 case of hyperbolic RN black holes.

\begin{figure}[ht]
\includegraphics[totalheight=1.7in]{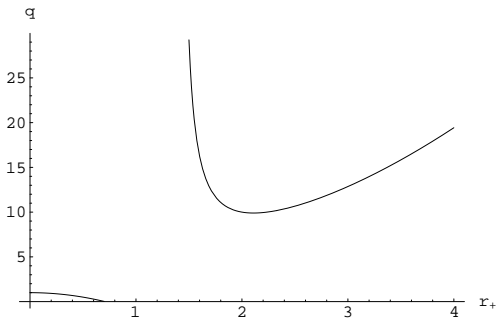}
 \caption{The curves denote the charge $ q/l^2$ versus the
 horizon radius $r_+/l$. The curve  starting from $q=1$ and $r_+=0$ and ending at
 $q=0$ and $r_+/l=1/\sqrt{2}$ denotes extremal black holes with vanishing temperature.
 The above curve stands for black holes with zero free energy.}
\end{figure}
The local stability of black hole thermodynamics is determined by
heat capacity. The heat capacity with a fixed charge is
 \begin{equation}
 C_{\tilde q}= \frac{\pi V_3 T}{2G} \frac{l^2(r_+^2
 +q)(3r_+^2(r_+^2+q)-l^2(3r_+^2+4q))}{2r_+^4+5qr_+^2
 -l^2(r_+^2+6q)-q^2-l^4}.
 \end{equation}
 The heat capacity with a fixed potential turns out to be
 \begin{equation}
 C_{\phi}= \frac{\pi V_3 T}{2G} \frac{l^2r_+^2(r_+^2
 +q)(3r_+^2-3l^2-q)}{2r_+^4+qr_+^2-l^2(2q +r_+^2) -q^2 -l^4}.
 \end{equation}
In Fig.~3 and 4 we plot the regions where the heat capacities
 $C_{\tilde q}$ and $C_{\phi}$ are negative, respectively.
\begin{figure}[ht]
\includegraphics[totalheight=1.7in]{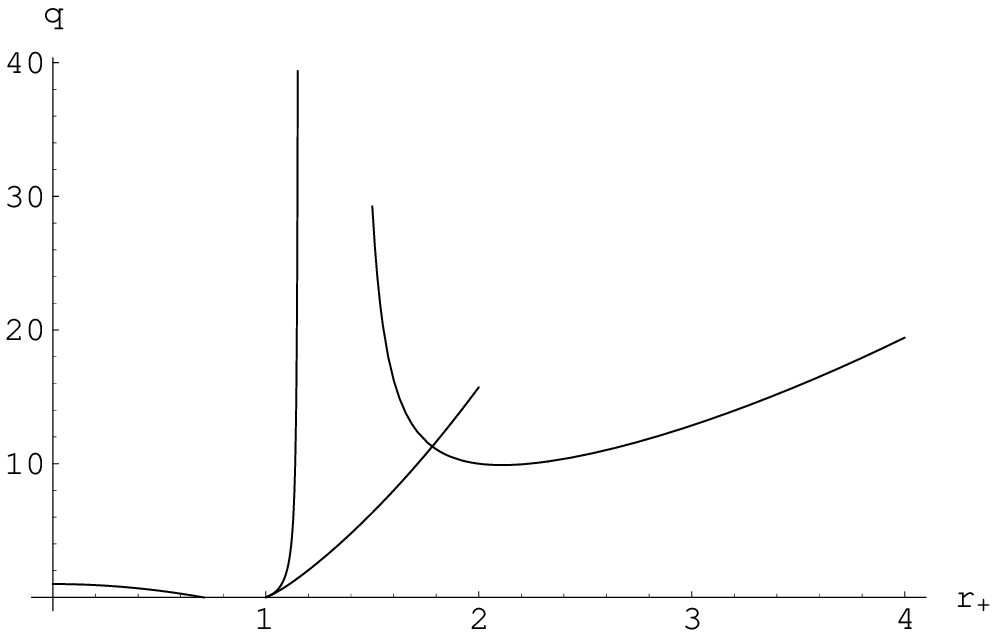}
 \caption{The curves denote the charge $ q/l^2$ versus the
 horizon radius $r_+/l$. The two curves starting  at $q=0$ and $r_+/l=1$
 enclose a region where the heat
 capacity with a fixed charge $C_{\tilde q}$ is negative. In other physical
 regions $C_{\tilde q}$ is positive.}
\end{figure}
\begin{figure}[ht]
\includegraphics[totalheight=1.7in]{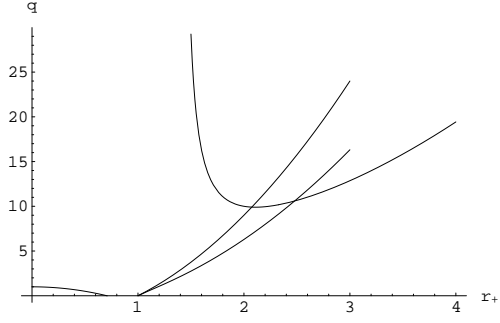}
 \caption{The curves denote the charge $ q/l^2$ versus the
 horizon radius $r_+/l$. The two  curves starting at $q=0$ and $r_+/l=1$
  enclose a region where the heat
 capacity with a fixed potential $C_{\phi}$  is negative. In other physical
 regions $C_{\phi}$ is positive.}
\end{figure}

\section{Hyperbolic black holes in $D=4$ gauged supergravity}

The black holes in D=4, N=8 gauged supergravity has been found in
\cite{DuffLiu,Sabra} (see also \cite{Duff}),
\begin{eqnarray}
&& ds^2 =-(H_1H_2H_3H_4)^{-1/2} fdt^2 +(H_1H_2H_3H_4)^{1/2}(fdr^2
+r^2 d\Omega_{2,k}^2), \nonumber \\
&& X_i = H_i^{-1}(H_1H_2H_3H_4)^{1/4}, \ \ \ A^i
=\sqrt{k}(1-H_i^{-1})\coth\beta_i dt,
\end{eqnarray}
where \begin{equation}
 f = k
-\frac{\mu}{r}+\frac{r^2}{l^2}H_1H_2H_3H_4, \ \ \ H_i = 1
+\frac{\mu\sinh^2\beta_i}{k r}, \ \ \ i = 1,2,3,4.
\end{equation}
As the case of D=5, the supersymmetric limit of the solution is
obtained by taking $\mu \to 0$ and $\beta_i \to \infty$ while
keeping $\mu \sinh^2\beta_i$ finite. But the supersymmetric limit
can be obtained only for the cases $k=1$ and $k=0$ (In the case of
$k=0$, one has to first take $\sinh^2\beta_i \to k\sinh^2\beta_i$
as the case of $D=5$). When $k=-1$, we have to make the
replacement $\beta_i \to \i \alpha_i$, and then obtain a real
gauge potential. In this case, one has $q_i = \mu\sin^2\alpha_i$
and the physical charge $\tilde q_i =\mu \sin\alpha_i
\cos\alpha_i$. Once again, $\mu$ and $\alpha_i$ are not good
parameters because of $0\le |q_i| \le |\mu |$. As the case of
$D=5$, when the solution has only a charge parameter (which
includes four cases: all four charges are equal; three charges are
equal and  one vanishes; two charges are equal and the other
two vanish; and one of charges does not vanishes and the other three
charges are zero), we can take $m$ and $\beta$ as two good
parameters: $q= m\sinh^2\beta$ and $\mu = m +2q$. And in this
case, one has the physical charge $\tilde q^2 = q
(m+q)=m^2\sinh^2\beta \cosh^2\beta$. In this way, $m$ and $q$ can
vary independently in the range $(-\infty,\infty)$.

The horizon radius of the hyperbolic black hole is determined by
the equation
\begin{equation}
\mu = r_+(-1 +\frac{1}{r_+^2 l^2}\Pi_i \rho_i), \ \ \  \rho_i =
r_+ +q_i.
\end{equation}
For the black hole, the associated  mass $M$, Hawking temperature
$T$, entropy $S$, and the chemical potential $\phi_i$  conjugate
to the physical electric charge $\tilde q_i$ are
 \begin{eqnarray}
 && M = \frac{V_2}{16\pi G}\left(2\mu -\sum_i q_i\right), \\
 && T =\frac{r_+}{4\pi\sqrt{\Pi_i\rho_i}}\left(-1
 -\frac{\Pi_i\rho_i}{l^2r_+^2}
 +\frac{\Pi_i\rho_i}{l^2r_+}\sum_i\frac{1}{\rho_i}\right), \\
&& S= \frac{V_2}{4G}\sqrt{\Pi_i\rho_i},\\
&& \phi_i=\frac{V_2}{16\pi G}\frac{\tilde q_i}{\rho_i},
 \end{eqnarray}
respectively. They obey $dM = TdS +\sum_i \phi_i d\tilde q_i$. The
Gibbs free energy reads
\begin{equation}
{\cal G} =\frac{V_2}{16\pi G}\left (-r_+
-\frac{\Pi_i\rho_i}{l^2r_+}\right),
\end{equation}
which is always negative, while the Helmholtz free energy of the
hyperbolic black hole is
\begin{equation}
\label{4eq9}
{\cal F}= \frac{V_2}{16\pi G} \left(-r_+
-\frac{\Pi_i\rho_i}{l^2r_+}-\sum_iq_i
+\frac{\Pi_i\rho_i}{l^2r_+}\sum_i\frac{q_i}{\rho_i}\right).
\end{equation}
As the case of $D=5$, the reference background is chosen as the
``zero mass" black hole with $\mu=q_i=0$. The discussions of the
reference background dependence of the free energies for the
hyperbolic RN black holes are of course still valid for the D=4
hyperbolic black holes. We do not repeat here.

For a single charge case, the free energy reduces to
\begin{equation}
{\cal F}=\frac{ V_2}{16\pi G} \left(-r_+ -\frac{r_+^3}{l^2}-q
\right).
\end{equation}
The heat capacities are
\begin{equation}
C_{\tilde q} = \frac{\pi V_2T
}{G}\frac{l^2r_+^2(r_++q)(2r_+^2(r_++q)-l^2(2r_++3q))}{3r_+^4(r_++2q)
  -r_+l^2(2q^2+7qr_++2r_+^2)-l^4(r_++q)},
  \end{equation}
  for a given electric charge,  and
\begin{equation}
C_{\phi}= \frac{\pi V_2 T}{G} \frac{l^2r_+^2 (r_++q)
((2r_+-q)r_+-2l^2))}{r_+^2(3r_+^2+qr_+-2q^2)-l^2r_+(3q+2r_+)-l^4},
\end{equation}
for a given gauge potential,  where the Hawking temperature is
\begin{equation}
T =\frac{1}{4\pi \sqrt{r_+\rho}}\left(-1 +\frac{2r_+ \rho}{l^2}
  +\frac{r_+^2}{l^2}\right),  \ \ \ \rho= r_+ +q.
  \end{equation}
We see that for a single charge case (say, $q_1=q$ and
$q_2=q_3=q_4=0$), the Helmholtz free energy is also always
negative. But to keep the positive definiteness of the Hawking
temperature, one has $q > l^2 (1-3r^2_+/l^2)/2r_+^2 $ when $0
<r_+^2/l^2 <1/3$ (while the temperature is always of positive
definiteness when $r_+^2/l^2 >1/3$). Therefore the Hawking-Page
phase transition will not occur in this case. The black hole phase
is always globally dominant in both the grand canonical
ensemble and the canonical ensemble. However, the hyperbolic
black holes are not always locally thermodynamically stable. In
Fig.~5 and 6 the thermodynamically unstable regions are plotted in
the canonical ensemble and grand canonical ensemble, respectively.
In these two figures, the curve starting from $q\to \infty$ and
$r_+=0$ and ending at $q=0$ and $r_+/l=1/\sqrt{3}\approx 0.58$
denotes extremal black holes, the region below that curve should
be excluded in the physical phase space.

\begin{figure}[ht]
\includegraphics[totalheight=1.7in]{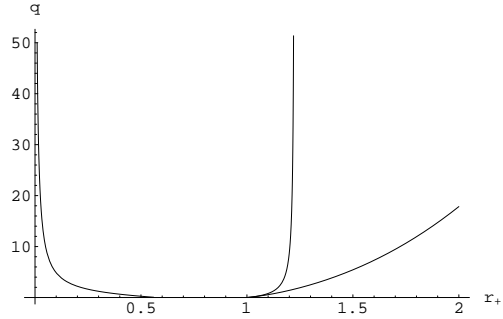}
 \caption{The curves denote the charge  $ q/l$ versus the
 horizon radius $r_+/l$. The first curve from left to right denotes extremal black holes with
 vanishing temperature. The two curves starting at $q=0$ and $r_+/l=1$ enclose
 a region where $C_{\tilde q} <0$. In other physical regions $C_{\tilde q}>0$.}
\end{figure}
\begin{figure}[ht]
\includegraphics[totalheight=1.7in]{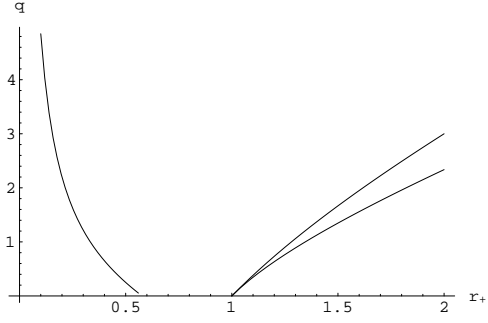}
 \caption{The curves denote the charge $ q/l$ versus the
 horizon radius $r_+/l$.  The first curve from left to right denotes extremal black holes with
 vanishing temperature. The region enclosed by the two curves starting
 at $q=0$ and $r_+/l=1$ denotes black holes having negative heat capacity with a fixed
 potential. In other physical regions $C_{\phi}>0$.}
\end{figure}

\section{Hyperbolic black holes in $D=7$ gauged supergravity}

The black hole solution in D=7, N=4 gauged supergravity
is~\cite{Gubser1} (see also \cite{Duff})
\begin{eqnarray}
&& ds^2= -(H_1H_2)^{-4/5}f dt^2 +(H_1H_2)^{1/5}(fdr^2 +r^2
d\Omega_{5,k}^2), \nonumber \\
&& X_i =H_i^{-1}(H_1H_2)^{2/5}, \ \ \ A^i =\sqrt{k}(1-H_i^{-1})dt,
\end{eqnarray}
where
\begin{equation}
f = k-\frac{\mu}{r^4} +\frac{r^2}{l^2}H_1H_2, \ \ \ H_i =
1+\frac{\mu \sinh^2\beta_i}{kr^4}, \ \ \ i=1,2.
\end{equation}
As the cases of $D=5$ and $D=4$, when $k=-1$, in order to get a
real gauge potential, one can take $\beta_i \to  \i \alpha_i$. In
this case, $q_i = \mu \sin^2\alpha_i$ and $\tilde q_i^2 =
\mu^2\sin^2\alpha_i \cos^2\alpha_i$. Therefore $\mu$ and
$\alpha_i$ are not good parameters as the cases of $D=5$ and
$D=4$. When the solution has only one charge parameter, once again,
we can remedy this difficulty by taking $m$ and $\beta$ as
two parameters describing the solution, $q= m \sinh^2\beta$ and
$\mu = m +2 q$, and then the physical charge $\tilde q^2 = m^2
\sinh^2 \beta \cosh^2\beta$. The single charge parameter case
includes two cases: (1) the two charges are equal,
$q_i=q_2=q$; (2) one of them vanishes, say $q_1=q$ and $q_2=0$.

 For this hyperbolic black hole, we obtain the associated mass,
 Hawking temperature, entropy and chemical
potential
\begin{eqnarray}
&& M = \frac{V_5}{4\pi G}\left(\frac{5}{4}\mu -\sum_i q_i\right),
\\
&& T =\frac{r_+^3}{\pi \Pi_i\rho_i^2}\left(-1
-\frac{\Pi_i\rho_i^4}{2l^2r_+^6}
   +\frac{\Pi_i\rho_i^4}{l^2r_+^2}\sum_i\frac{1}{\rho_i^4}\right),
    \\
&& S= \frac{V_5}{4G}r_+\Pi_i\rho_i^2, \\
&& \phi_i = \frac{V_5}{4\pi G} \frac{\tilde q_i}{\rho_i^4},
\end{eqnarray}
where the horizon is determined by the equation
\begin{equation}
\mu = r_+^4\left(-1 +\frac{1}{l^2r_+^6}\Pi_i\rho_i^4\right), \ \ \
\rho_i^4 = r^4_+ +q_i.
\end{equation}
It is easy to show $dM =TdS +\sum_i \phi_i \tilde q_i$.  The Gibbs
free energy is found to be
\begin{equation}
{\cal G}= \frac{V_5}{16\pi G} \left(-r_+^4
-\frac{\Pi_i\rho^4_i}{l^2r_+^2}\right),
\end{equation}
and the Helmholtz free energy
\begin{equation}
{\cal F}= \frac{V_5}{16\pi G}\left(-r_+^4-4 \sum_iq_i
-\frac{4r_+^2}{l^2}\Pi_i\rho_i^4 \sum_i\frac{1}{\rho_i^4}
  +7\frac{\Pi_i\rho_i^4}{l^2r_+62}\right).
  \end{equation}
Once again, we see that the Gibbs free energy is always negative.
As a result the Hawking-Page phase transition will not happen in
the grand canonical ensemble, while in the canonical ensemble it
is possible. Of course, the free energies and the Euclidean
actions depend on the choice of reference background. Here as the
case of hyperbolic RN black hole, a seven dimensional ``zero mass"
black hole has been taken. Similar discussions made for the
hyperbolic RN black holes can be repeated. But we do not present
them them here.

For the  single charge case, where one of charges vanishes, we
have
\begin{equation}
T = \frac{r_+}{\pi \rho^2}\left(-1
+\frac{\rho^4}{2l^2r_+^2}+\frac{r_+^2}{l^2}\right), \ \ \ \rho^4
=r_+^4 +q,
\end{equation}
and the free energy reduces to
\begin{equation}
{\cal F} = \frac{V_5}{16\pi G}\left(-r_+^4 -4 q
-\frac{r_+^6}{l^2}+3\frac{r_+^2}{l^2}q\right).
\end{equation}
In order to have ${\cal F} \le 0$, one has to have $q <
r_+^4(1+r^2_+/l^2)/(3r^2_+/l^2-4)$ for $r_+^2/l^2 >4/3$. When
$r_+^2/l^2 <4/3$, the free energy is always negative. On the other
hand, the positive definiteness of the temperature has to be
guaranteed. Therefore $q >r_+^2 l^2 (2-3r_+^2/l^2)$ when $0 \le
r_+/l < \sqrt{6}/3 \approx 0.82$. In Fig.~7 the curve standing for
the extremal black holes and the curve having zero free energy are
plotted. The region below the extremal black hole curve starting
from $q=0$ and $r_+=0$ and ending at $q=0$ and $r_+/l =0.82$
should be excluded in the physical phase space. Note that for the
sake of demonstration, the charge $q$ for extremal black holes is
amplified by ten times in Fig.~7 and 8 and 9.

\begin{figure}[ht]
\includegraphics[totalheight=1.7in]{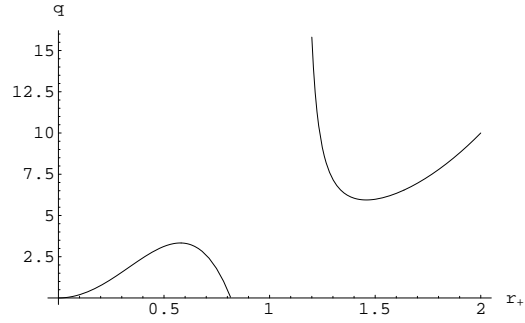}
 \caption{The curves denote the charge  $ q/l^4$ versus the
 horizon radius $r_+/l$. The curves starting at $q=0$ and $r_+=0$ and ending
 at $q=0$ and $r_+/l=0.82$ denotes extremal black holes with vanishing Hawking
 temperature. The another curve stands for the black holes having zero Helmholtz
 free energy, in the region below this curve black holes have negative free energy.
 Note that for the sake of demonstration, the charge $q$ for extremal black holes is
 amplified by ten times here.}
\end{figure}
The heat capacities are
\begin{equation}
C_{\tilde q}=\frac{\pi
V_5T}{4G}\frac{2l^2r_+^4(r_+^4+q)(5r_+^2(r_+^4+q)-l^2(5r_+^4+6q))}
    {3r_+^2(r_+^8+4qr_+^4-q^2)-l^2(-2q^2+17qr_+^4+r_+^8)-l^4(2r_+^6-4qr_+^2)}
\end{equation}
for a fixed charge, and
\begin{equation}
C_{\phi}= \frac{\pi
V_5T}{4G}\frac{2l^2r_+^4(5r_+^8-5l^2r_+^6+4qr_+^4-5l^2qr_+^2-q^2)}
   {3r_+^8 -l^2r_+^6+2qr_+^4-2l^4r_+^4-3l^2qr_+^2-q^2},
   \end{equation}
for a fixed gauge potential, respectively. In Fig.~8, the region
for the locally thermodynamically unstable black holes in the
grand canonical ensemble is plotted. The narrow region between
the two curves both starting at $q=0$ and $r_+/l=1$ has $C_{\phi}<0$.

\begin{figure}[ht]
\includegraphics[totalheight=1.7in]{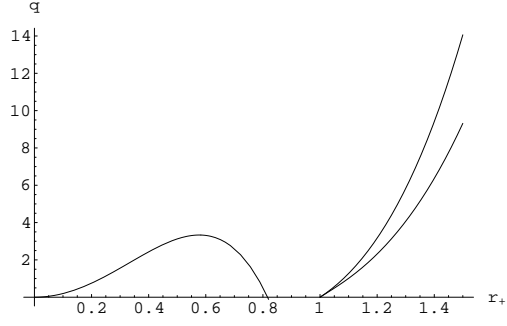}
 \caption{The curves denote the charge  $ q/l^4$ versus the
 horizon radius $r_+/l$.  In the region between the two curves starting at $q=0$ and
 $r_+/l=1$ one has $C_{\phi}<0$. In the other physical region, $C_{\phi}>0$.}
\end{figure}

On the other hand, in the canonical ensemble the hyperbolic black
hole is locally thermodynamical stable in the regions as follows:
(1) When $0<r_+/l <\sqrt{2/3}\approx 0.82$,
\begin{equation}
  \left (2-3\frac{r^2_+}{l^2}\right)\frac{r_+^2}{l^2} < \frac{q }{l^4}
  <
  \frac{q_{c1}}{l^4}
\end{equation}
where
\begin{equation}
 \frac{q_{c1}}{l^4}=\frac{17 r_+^4/l^4 -12
r_+^6/l^6 -4r_+^2/l^2
  +\sqrt{(12r_+^6/l^6-17r_+^4/l^4-4r_+^2/l^2)^2 -4
  (2-3r_+^2/l^2)(3r_+^{10}/l^{10}-r_+^8/l^8-2r_+^6/l^6)}}{2(2-3r_+^2/l^2)}.
  \end{equation}
(2) When $ \sqrt{2/3} <r_+/ l < 1 $, the black hole is always
thermodynamical stable for any $q$. (3) When $1 < r_+/l <
\sqrt{6/5}\approx 1.1$, the black hole is locally thermodynamically
 stable for  $q>q_{2c}$ or $q<q_{3c}$. (4) When
$r_+/l>\sqrt{6/5}$, one has to have $q <q_{3c}$, where
\begin{equation}
 \frac{q_{2c}}{l^4} = \frac{5r_+^4(r_+^2/l^2)}{l^4(6-5r_+^2/l^2)}
\end{equation}
and
\begin{equation} \frac{q_{c3}}{l^4}=\frac{17 r_+^4/l^4 -12
r_+^6/l^6 -4r_+^2/l^2
  -\sqrt{(12r_+^6/l^6-17r_+^4/l^4-4r_+^2/l^2)^2 -4
  (2-3r_+^2/l^2)(3r_+^{10}/l^{10}-r_+^8/l^8-2r_+^6/l^6)}}{2(2-3r_+^2/l^2)}.
  \end{equation}

The thermodynamical stable regions are plotted in Fig.~9. The
region enclosed by the two curves starting from $q=0$ and
$r_+/l=1$ are thermodynamic unstable with a negative heat capacity
for a fixed charge $C_{\tilde q}$. The another thermodynamic
unstable region is enclosed by the charge axe $q$, the extremal
black hole curve and the one starting from $q=0$ and $r_+=0$ and
going to infinity at $r_+/l= \sqrt{2/3}$.  In other regions the hyperbolic
black hole is locally thermodynamically stable with a positive heat capacity
$C_{\tilde q}$.

\begin{figure}[ht]
\includegraphics[totalheight=1.7in]{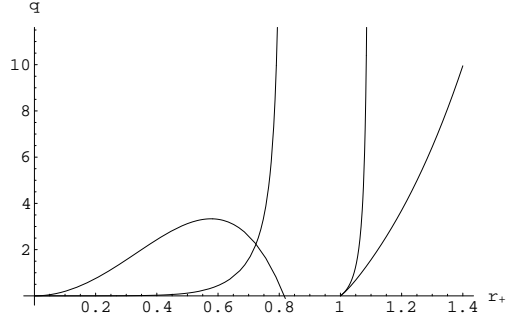}
 \caption{The curves denote the charge  $ q/l^4$ versus the
 horizon radius $r_+/l$. In the
region enclosed by the two curves starting at $q=0$ and
$r_+/l=1$ the black hole is thermodynamically unstable with a
negative heat capacity $C_{\tilde q}$. The black hole is also locally
unstable in the region which is enclosed by
the charge axe $q$, the extremal black hole curve and the one
starting at $q=0$ and $r_+=0$ and going to infinity at
$r_+/l=\sqrt{2/3}$.}
\end{figure}

\section{Conclusions and Discussions}
We discussed the thermodynamics, global and local stability of hyperbolic charged black holes
in AdS spaces. These black holes include hyperbolic RN black holes in arbitrary dimensions,
hyperbolic charged black holes in D=5, 4 and 7 dimensional gauged supergravities. In particular, we
emphasized how to choose appropriate parameters to parameterize the solution in the isotropic
coordinates so that the chosen parameters can cover the whole phase space.
We found that the entropy of the hyperbolic RN black holes can be expressed
by a Cardy-Verlinde-like formula, but the Casimir energy is negative, which generalized the
discussion for the neutral black holes to the charged case.  It is easy to check that similar
formula holds for the hyperbolic black holes in D=5, 4 and 7 dimensional gauged supergravities.

For the hyperbolic RN black holes in arbitrary dimensions, we found that the Gibbs free energy is
always negative and heat capacity with a fixed gauge potential is always positive. This shows
that in the grand canonical ensemble, the black hole is stable not only globally but also
locally; the Hawking-Page phase transition will not happen and black hole phase is dominant in the
dual CFTs. In the canonical ensemble, the heat capacity with a given electric charge is still positive,
indicating the local stability of the black hole, but the Helmholtz free energy can change its
sign. This implies that a Hawking-Page phase transition will occur in the canonical ensemble.

For the hyperbolic black holes in D=5, 4 and 7 dimensional gauged supergravities, the negative
definiteness of Gibbs free energy still persists, but the heat capacity with a given gauge potential
could change its sign. This implies that in the grand canonical ensemble, the hyperbolic black
holes are globally stable, but not locally. The black hole phase is globally preferred and
dominant in the dual CFTs, but the black holes cannot always be thermal equilibrium with an
infinite thermal bath.  In the canonical ensemble, there is always a region where the heat capacity
with a given electric charge is negative, indicating the local instability of black holes. The
Helmholtz free energy for a single charge black holes can change its sign, except for the case
of $D=4$. This implies that a Hawking-page phase transition will appear as the case of hyperbolic
RN black holes.  We noted that the Helmholtz free energy  is always negative for a single charged
black hole in D=4 gauged supergravity. To understand this situation, it is helpful to
recall the case of rotating M2-branes~\cite{CS2,Gubser1,Gubser2}, there in the case of a
single rotation
parameter, the free energy is also always negative, but it will change its sign for multiple rotation
parameters. According to the relationship between the black hole solutions in D=4 gauged
supergravity and the rotating M2-branes~\cite{Gubser1,CEJM}) (also see \cite{Duff}), we expect
that the Helmholtz free energy will also change its sign for multiple charged hyperbolic black holes
in D=4 gauged supergravity, which in fact is already predicted by (\ref{4eq9}).

We point out here that when analyzing the stability and phase structure of a single charged black
holes in D=5, 4 and 7 gauged supergravities, we restricted ourselves to the case $q>0$. As we
discussed in the text, for the hyperbolic black holes, it is also possible to have $q<0$,
and in this case the solution still has a black hole structure. Therefore it would be of interest
to extend the discussion to the case of $q<0$, and also to the case of multiple charges.
Further, the implication in the AdS/CFT correspondence is worth investigating~\cite{Min}.

When discussing the global stability of AdS black holes, one has to calculate the Euclidean action
associated with the black hole. Due to the infinite volume, the Hilbert-Einstein action for the
black hole configuration is divergent. To get a finite Euclidean action,
 usually one can use the so-called background subtraction
approach~\cite{BJ}, or the surface counterterm approach~\cite{sca}. In the background subtraction
approach, one has to choose an appropriate background: for asymptotically flat spaces, the
 Minkowski space is a suitable one~\cite{York2}, while for asymptotically AdS
 spaces, the AdS space is a suitable choice. But in some cases,
 for example, for the Taub-NUT and Taub-Bolt spaces, an appropriate vacuum background
 is difficult to choose. For those cases, the surface counterterm
 approach works very well~\cite{sca}. In this paper, we have used
 the background substraction approach~\cite{BJ}.  The chosen
 background is the AdS vacuum solution with vanishing mass
 parameter and charge parameter. When calculating the gravitational mass of
 the hyperbolic black hole, we used the formula (\ref{3eq12}), which
 is derived from the Euclidean action of black holes by subtracting the
 contribution of chosen vacuum background~\cite{BJ}.  Through such defined
 mass,  the resulting free energies associated with black holes have the relation to the
 Euclidean action of black holes as ${\cal G}=T {\cal I}_g$ and
 ${\cal F}= T {\cal I}_c$ in the grand canonical ensemble and
 canonical ensemble~\cite{York}, respectively.  Here the Euclidean
 actions ${\cal I}_g$ and ${\cal I}_c$ are understood as the differences between
 the action of black holes and the one for the vacuum background (see for example
  \cite{York,York2}). Therefore, the Euclidean action and free
  energy have the same signs and then the global stability of
  AdS black holes can be inferred from the behavior of free
  energy in the grand canonical ensemble and canonical ensemble.
  For the equivalence, one may further refer to
  \cite{Witten,Birm,Gubser1}. For the hyperbolic AdS black holes, the choice of the
  vacuum background is not unique as we pointed out before.  One of
  the choices is to take the extremal black hole as the background
  (see for example \cite{Birm}). Another natural choice is just to
  take the AdS background without any excitation, namely black
  hole solutions with vanishing mass parameter (we chose this
  background). The surface counterterm approach reveals that
  both the vacuum backgrounds might be suitable~\cite{Emp}.
  Anyway, it is an interest issue to further study the dependence
  of global stability on the choice of vacuum background. In
  particular, it is of great importance to investigate whether or
  not there is a mechanism to uniquely determine a physical vacuum
  background via the AdS/CFT correspondence.

  Finally we stress that when the heat capacities are negative,
  both the grand canonical ensemble and canonical ensemble are not
  well-defined.  In that case, one way to remedy this difficulty
  is to put the AdS black hole in a finite cavity as the case of a
  Schwarzschild black hole in a cavity (see the paper by York in
  \cite{York2}). In this paper, we are also interested in thermodynamics and
  stability of dual CFTs which reside on the boundary of AdS space, so the
  cavity is put on the boundary of AdS space. For large AdS Schwarzschild black
  holes~\cite{HP,Witten}, the heat capacity is positive. Therefore
  it is well-defined to discuss thermodynamics and stability of
  the black hole in the canonical ensemble. For small AdS
  Schwarzschild black holes, the heat capacity is negative. This of
  course implies that one cannot use the canonical ensemble to
  study the small AdS Schwarzschild black hole. Instead it is
  understood that this small black hole with negative heat
  capacity cannot be in thermal equilibrium with an infinite heat
  bath, as the case of Schwarzschild black hole. In the AdS/CFT
  correspondence, on the side of dual CFTs, the phase of small black
  hole should be replaced by a thermal AdS background, which has
  a positive heat capacity~\cite{HP,Witten}. In our case, the
  situation is similar to the case of AdS Schwarzschild black hole.
  The hyperbolic charged black holes with heat capacity cannot be
  in thermal equilibrium with an infinite heat bath surrounding the
  black hole. On the side of dual CFTs, the corresponding phase is not
  controlled by these black holes with negative heat capacity,
  but by other gravity configurations, for example a thermal AdS background
  with a fixed gauged potential.  No doubt it is of interest to
  study thermodynamics and stability of these hyperbolic charged black
  holes in a finite cavity as the cases of charged AdS black holes
  with positive constant curvature horizon~\cite{BLP,RN} and with
  zero curvature horizon~\cite{Lemos}. In particular, one should
  be quite interested in whether or not the thermodynamics
  of AdS black hole in a cavity has a dual description on the side
  of dual CFTs.

\section*{Acknowledgments} RGC would like to thank J.X. Lu and
 W. Sabra for helpful
  discussions at the initial stage of this work.
  He also would like to express his gratitude to the Physics
  Department, Baylor University for its hospitality. This work was
  supported by Baylor University, a grant from Chinese Academy of
  Sciences, a grant from NSFC, China (No. 10325525), and a grant
   from the Ministry of Science and Technology of China (No.
   TG1999075401).


\begin{references}
\bibitem{FSW}J.~L.~Friedman, K.~Schleich and D.~M.~Witt,
             Phys.\ Rev.\ Lett.\  {\bf 71}, 1486 (1993)
             [Erratum-ibid.\  {\bf 75}, 1872 (1995)]
             [arXiv:gr-qc/9305017];
T.~Jacobson and S.~Venkataramani,
Class.\ Quant.\ Grav.\  {\bf 12}, 1055 (1995)
[arXiv:gr-qc/9410023].

\bibitem{Topology} J.~P.~S.~Lemos,
                   Phys.\ Lett.\ B {\bf 353}, 46 (1995) [arXiv:gr-qc/9404041];
                   J.~P.~Lemos,
                   Class.\ Quant.\ Grav.\  {\bf 12}, 1081 (1995) [gr-qc/9407024];
                   J.~P.~S.~Lemos and V.~T.~Zanchin,
                   Phys.\ Rev.\ D {\bf 54}, 3840 (1996) [arXiv:hep-th/9511188].
                   C.~G.~Huang and C.~B.~Liang,
                   Phys.\ Lett.\ A {\bf 201} (1995) 27;
                   R.~G.~Cai and Y.~Z.~Zhang,
                   Phys.\ Rev.\ D {\bf 54}, 4891 (1996) [arXiv:gr-qc/9609065];
                   R.~G.~Cai, J.~Y.~Ji and K.~S.~Soh,
                   Phys.\ Rev.\ D {\bf 57}, 6547 (1998) [arXiv:gr-qc/9708063].
                   R.~G.~Cai,
                   Nucl.\ Phys.\ B {\bf 524}, 639 (1998) [arXiv:gr-qc/9801098];
                   S.~Aminneborg, I.~Bengtsson, S.~Holst and P.~Peldan,
                   Class.\ Quant.\ Grav.\  {\bf 13}, 2707 (1996) [gr-qc/9604005].
                   R.~B.~Mann,
                   Class.\ Quant.\ Grav.\  {\bf 14}, L109 (1997) [gr-qc/9607071];
                   R.~B.~Mann,
                   Nucl.\ Phys.\ B {\bf 516}, 357 (1998) [hep-th/9705223].
                   L.~Vanzo,
                   Phys.\ Rev.\ D {\bf 56}, 6475 (1997) [gr-qc/9705004].
                   D.~Klemm,
                   Class.\ Quant.\ Grav.\  {\bf 15}, 3195 (1998) [gr-qc/9808051];
                   D.~Klemm, V.~Moretti and L.~Vanzo,
                   Phys.\ Rev.\ D {\bf 57}, 6127 (1998) [Erratum-ibid.\ D {\bf 60},
                   109902 (1998)] [gr-qc/9710123];
                   M.~Banados, A.~Gomberoff and C.~Martinez,
                   Class.\ Quant.\ Grav.\  {\bf 15}, 3575 (1998) [hep-th/9805087].
                   M.~F.~A.~da Silva, A.~Wang, F.~M.~Paiva and N.~O.~Santos,
                   Phys.\ Rev.\ D {\bf 61}, 044003 (2000) [arXiv:gr-qc/9911013];
                   R.~G.~Cai,
                   Phys.\ Lett.\ B {\bf 572}, 75 (2003)
                   [arXiv:hep-th/0306140];
R.~Aros, R.~Troncoso and J.~Zanelli,
Phys.\ Rev.\ D {\bf 63}, 084015 (2001) [hep-th/0011097].
R.~G.~Cai,
Phys.\ Rev.\ D {\bf 65}, 084014 (2002) [arXiv:hep-th/0109133].
R.~G.~Cai,
Phys.\ Lett.\ B {\bf 582}, 237 (2004) [arXiv:hep-th/0311240];
W.~L.~Smith and R.~B.~Mann,
Phys.\ Rev.\ D {\bf 56}, 4942 (1997) [gr-qc/9703007];
Y.~Wu, M.~F.~A.~da Silva, N.~O.~Santos and A.~Wang,
Phys.\ Rev.\ D {\bf 68}, 084012 (2003) [arXiv:gr-qc/0309002];
S.~Nojiri and S.~D.~Odintsov,
Phys.\ Lett.\ B {\bf 521}, 87 (2001) [Erratum-ibid.\ B {\bf 542},
301 (2002)] [arXiv:hep-th/0109122];
M.~Cvetic, S.~Nojiri and S.~D.~Odintsov,
Nucl.\ Phys.\ B {\bf 628}, 295 (2002) [arXiv:hep-th/0112045].
S.~Nojiri and S.~D.~Odintsov,
Phys.\ Rev.\ D {\bf 66}, 044012 (2002) [arXiv:hep-th/0204112].
Y.~M.~Cho and I.~P.~Neupane,
Phys.\ Rev.\ D {\bf 66}, 024044 (2002) [arXiv:hep-th/0202140].
I.~P.~Neupane,
Phys.\ Rev.\ D {\bf 67}, 061501 (2003) [arXiv:hep-th/0212092].
I.~P.~Neupane,
arXiv:hep-th/0302132;
and references therein.
 \bibitem{BLP}D.~R.~Brill, J.~Louko and P.~Peldan,
               Phys.\ Rev.\ D {\bf 56}, 3600 (1997) [arXiv:gr-qc/9705012];
 \bibitem{Birm}D.~Birmingham,
               Class.\ Quant.\ Grav.\  {\bf 16}, 1197 (1999) [hep-th/9808032].
 \bibitem{CS} R.~Cai and K.~Soh,
              Phys.\ Rev.\ D {\bf 59}, 044013 (1999) [gr-qc/9808067];
\bibitem{CEJM}A.~Chamblin, R.~Emparan, C.~V.~Johnson and R.~C.~Myers,
              Phys.\ Rev.\ D {\bf 60}, 064018 (1999) [arXiv:hep-th/9902170];

\bibitem{Klemm}M.~M.~Caldarelli and D.~Klemm,
Nucl.\ Phys.\ B {\bf 545}, 434 (1999) [arXiv:hep-th/9808097].

\bibitem{CV}R.~G.~Cai,
Phys.\ Rev.\ D {\bf 63}, 124018 (2001) [arXiv:hep-th/0102113].

\bibitem{HP}S.~W.~Hawking and D.~N.~Page,
            Commun.\ Math.\ Phys.\  {\bf 87}, 577 (1983).
\bibitem{AdS}J.~Maldacena,
             Adv.\ Theor.\ Math.\ Phys.\  {\bf 2}, 231 (1998) [Int.\ J.\
             Theor.\ Phys.\  {\bf 38}, 1113 (1998)] [hep-th/9711200];
              S.~S.~Gubser, I.~R.~Klebanov and A.~M.~Polyakov,
             Phys.\ Lett.\ B {\bf 428}, 105 (1998) [hep-th/9802109];
              E.~Witten,
             Adv.\ Theor.\ Math.\ Phys.\  {\bf 2}, 253 (1998) [hep-th/9802150].
\bibitem{Witten}E.~Witten,
                Adv.\ Theor.\ Math.\ Phys.\  {\bf 2}, 505 (1998)
                [arXiv:hep-th/9803131].

\bibitem{Gubser} S.~S.~Gubser,
                 Nucl.\ Phys.\ B {\bf 551}, 667 (1999)
                 [arXiv:hep-th/9810225].

\bibitem{CS2}R.~G.~Cai and K.~S.~Soh,
             Mod.\ Phys.\ Lett.\ A {\bf 14}, 1895 (1999)
             [arXiv:hep-th/9812121].

\bibitem{Gubser1}M.~Cvetic and S.~S.~Gubser,
JHEP {\bf 9904}, 024 (1999) [arXiv:hep-th/9902195].

\bibitem{Gubser2}M.~Cvetic and S.~S.~Gubser,
                 JHEP {\bf 9907}, 010 (1999)
                 [arXiv:hep-th/9903132].

\bibitem{Harm}T.~Harmark and N.~A.~Obers,
              JHEP {\bf 0001}, 008 (2000)
              [arXiv:hep-th/9910036].

\bibitem{York}H.~W.~Braden, J.~D.~Brown, B.~F.~Whiting and J.~W.~.~York,
Phys.\ Rev.\ D {\bf 42}, 3376 (1990).

\bibitem{BCS}K.~Behrndt, M.~Cvetic and W.~A.~Sabra,
Nucl.\ Phys.\ B {\bf 553}, 317 (1999) [arXiv:hep-th/9810227].

\bibitem{Duff}M.~Cvetic {\it et al.},
Nucl.\ Phys.\ B {\bf 558}, 96 (1999) [arXiv:hep-th/9903214];
M.~J.~Duff,
arXiv:hep-th/9912164.

\bibitem{BJ}J.~D.~Brown and J.~W.~.~York,
Phys.\ Rev.\ D {\bf 47}, 1407 (1993);
S.~W.~Hawking and G.~T.~Horowitz,
Class.\ Quant.\ Grav.\  {\bf 13}, 1487 (1996)
[arXiv:gr-qc/9501014];
J.~D.~Brown, J.~Creighton and R.~B.~Mann,
Phys.\ Rev.\ D {\bf 50}, 6394 (1994) [arXiv:gr-qc/9405007].

\bibitem{DuffLiu}M.~J.~Duff and J.~T.~Liu,
Nucl.\ Phys.\ B {\bf 554}, 237 (1999) [arXiv:hep-th/9901149].

\bibitem{Sabra}W.~A.~Sabra,
Phys.\ Lett.\ B {\bf 458}, 36 (1999) [arXiv:hep-th/9903143].

\bibitem{Min} B.~McInnes,
              Nucl.\ Phys.\ B {\bf 660}, 373 (2003)
              [arXiv:hep-th/0205103].

\bibitem{sca}V.~Balasubramanian and P.~Kraus,
              Commun.\ Math.\ Phys.\  {\bf 208}, 413 (1999)
              [arXiv:hep-th/9902121];
R.~Emparan, C.~V.~Johnson and R.~C.~Myers,
Phys.\ Rev.\ D {\bf 60}, 104001 (1999)
[arXiv:hep-th/9903238];
P.~Kraus, F.~Larsen and R.~Siebelink,
Nucl.\ Phys.\ B {\bf 563}, 259 (1999)
[arXiv:hep-th/9906127].

\bibitem{York2} G.~W.~Gibbons and S.~W.~Hawking,
               Phys.\ Rev.\ D {\bf 15}, 2752 (1977);
               J.~W.~.~York,
Phys.\ Rev.\ D {\bf 33}, 2092 (1986).

\bibitem{Emp} R.~Emparan,
              JHEP {\bf 9906}, 036 (1999)
              [arXiv:hep-th/9906040].

\bibitem{RN}
C.~S.~Peca and J.~P.~S.~Lemos,
            Phys.\ Rev.\ D {\bf 59}, 124007 (1999)
            [arXiv:gr-qc/9805004].

\bibitem{Lemos} C.~S.~Peca and J.~P.~S.~Lemos,
                J.\ Math.\ Phys.\  {\bf 41}, 4783 (2000)
                [arXiv:gr-qc/9809029].

\end{references}
\end{document}